# Estimation of the yield curve for Costa Rica using combinatorial optimization metaheuristics applied to nonlinear regression

Andrés Quirós-Granados[1] and Javier Trejos-Zelaya[2][0000-0002-4459-9188]

[1] University of Costa Rica, San José 11501, Costa Rica
andquigr@gmail.com
[2] University of Costa Rica, San José 11501, Costa Rica
javier.trejos@ucr.ac.cr

**Abstract.** The term structure of interest rates or yield curve is a function relating the interest rate with its own term. Nonlinear regression models of Nelson-Siegel and Svensson were used to estimate the yield curve using a sample of historical data supplied by the National Stock Exchange of Costa Rica. The optimization problem involved in the estimation process of model parameters is addressed by the use of four well known combinatorial optimization metaheuristics: Ant colony optimization, Genetic algorithm, Particle swarm optimization and Simulated annealing. The aim of the study is to improve the local minima obtained by a classical quasi-Newton optimization method using a descent direction. Good results with at least two metaheuristics are achieved, Particle swarm optimization and Simulated annealing.

## 1      Introduction

The interest rate is essential in the modern economy, it refers to the payment of money from a debtor to a creditor by use of capital [12]. There are many factors that determine the level of interest rates: inflation risk, uncertainty, quality of information, random fluctuations and the period of investment, among others. Remaining constant all factors affecting the level of interest rates, except the period of investment is called term structure of rates interest [12].

In a technical document authored by Bank for International Settlements [2] presented methodologies and models used by 13 nations in the estimation of the yield curve, which highlight the parametric models of Nelson-Siegel and Svensson.

One of the papers, which is an important reference, is presented by the Central Bank of Canada [4]. The paper introduces the parametric models of Nelson-Siegel and Svensson for estimating the yield curve in the Central Bank of Canada. The optimization problem was faced with two methods called partial-estimation algorithm and full-estimation algorithm. It was concluded that the optimization process can be im-proved. Moreover, given the large size of the search space, genetic algorithms was suggested as a method that can improve the estimation.

The stock market in Costa Rica is small, therefore many of the existing methods are not feasible to implement. The paper of Barboza et al. [3] mentions that after reviewing existing models to estimate the curve, the most suitable for the Costa Rica market is the Svensson model. It also proposes a modification to the objective function, in order to consider topics such as historical data and volatility.

The optimization problem in the area of the yield curve for Costa Rica was worked by Piza et al. [23]. In that study numerical methods such as Gauss-Newton, gradient descent and Marquardt were used [6]. It was concluded that a successful optimization depends on the initial values and, also, indicated that only local minima were obtained. It is recommended the use of Metaheuristics to address the problem of finding the global minimum.

In the present paper, Metaheuristics are implemented to improve local minima that are achieved using methods that work with descent direction in the problem of estimating the parameters of the nonlinear regression models of Nelson-Siegel and Svensson.

The article is divided as follows: in Section 2 we present the data and it's particularities. Section 3 describes the yield curve models used, Nelson-Siegel and Svensson, and the optimization criterion to be minimized. In Section 4 are presented the heuristics we have used and the characteristics of their implementation. Results in the real Costarican data are contained in Section 5 and we conclude in Section 6, mainly that Simulated annealing and Particle swarms achieved better results.

## 2    Data

Historical data were provided by the Bolsa Nacional de Valores (BNV, Costa Rican-National Stock Exchange). These are bonds and zero-coupon bonds issued by the Central Bank and the Treasury Costa Rica, which are called *tp0*, *tp*, *bem0* and *bem*.

Data are for the period of February 23, 2015 to March 12, 2015, only emissions in colones, the Costa Rican national currency, were considered and there is not any restriction on the amounts of transactions.

Data were provided by the BNV and contain the following information:

- Description and acronym of the bond issuer.
- Classification of the instrument.
- Identification of the bond.
- Date of issuing.
- Date of expiration.
- Date of next coupon.
- ISIN code (for international identification).
- Currency.
- Periodicity: interval of times for coupon payments.
- Net rate: rate payed by the coupon issuer.

- Rate type: fixed, variable or without rate.

From the *book of closed operations*, we obtain the following information:
- Type of operation: in the primary or secondary market.
- Date of operation: day of the operation.
- Nominal yield: net yield obtained by the financial instrument in the operation.
- Price: price paid in the transaction, as well as
- Value of transaction in colones.

From the *book of buy and sell offers* we obtain the following information:
- Offer: quotation identifier.
- Facial amount: quotation amount.
- Yield: quote yield, net of tax.
- Price: proposed price in the quotation.
- Position: indicated whether it is a buy or a sell.

In the case that a financial instrument is present several times, we keep only the last appearance in the book of closed operations. From the books of buy and sell offers we calculate the average *bid-ask spread*; in some cases this spread cannot be calculated since there are only buy offers or only sell offers or there was no offer at all, in these cases the observation is not used.

Prices in these books are *clean prices*, for our estimation we use *dirty prices*, that is, the clean price added by cumulated interests.

The data base with 32 entries was reduced to 25 entries, after the elimination of observations that concentrated too much weight.

## 3   Yield curve estimation

The yield curve relates interest rates with its own term [10], [12], this rate is call spot interest rate.

The forward interest rate is an interest that is negotiated today for a transaction that will occur in the future [22]. The forward rate is an expectation of what the spot rate will be in the future [10].

If there are continuous rates $\delta_t$ and $\delta_s$ for terms $t$ and $s$, $(s < t)$, it is defined the forward continuous rate as [2], [3]:

$$f_{t,s} = \frac{t\delta_t - s\delta_s}{t - s}.$$

The instantaneous forward rate is obtained as a limit [3, 14, 22]:

$$f_t = \lim_{s \to t} f_{t,s}.$$

The Nelson-Siegel model [19], from 1987, proposes a continuous function to describe the shape of the instantaneous forward rate depending on the term $t$,

$$f_t = \beta_0 + \beta_1 e^{-\lambda t} + \beta_2 \lambda t e^{-\lambda t}. \quad (1)$$

From equation (1) a continuous function is obtained for the spot rate,

$$\delta_t = \beta_0 + \beta_1 \left(\frac{1 - e^{-\lambda t}}{\lambda t}\right) + \beta_2 \left(\frac{1 - e^{\lambda t}}{\lambda t} - e^{-\lambda t}\right).$$

The Svensson model (1994) [28] extends the Nelson-Siegel model by incorporating two parameters more: $\beta_3$ y $\lambda_2$. Thus, the continuous function for forward rate is,

$$f_t = \beta_0 + \beta_1 e^{-\lambda_1 t} + \beta_2 \lambda_1 t e^{-\lambda_1 t} + \beta_3 \lambda_2 t e^{-\lambda_2 t}, \quad (2)$$

and from (2) the function for the spot rate is

$$\delta_t = \beta_0 + \beta_1 \left(\frac{1 - e^{-\lambda_1 t}}{\lambda_1 t}\right) + \beta_2 \left(\frac{1 - e^{-\lambda_1 t}}{\lambda_1 t} - e^{\lambda_1 t}\right) + \beta_3 \left(\frac{1 - e^{-\lambda_2 t}}{\lambda_2 t} - e^{\lambda_2 t}\right).$$

If the spot rates for different maturities are available, the price of a bond can be calculated as [10]

$$Pr = \sum_{k=1}^{t} c \, e^{-\delta_k k} + F e^{-\delta_t t}, \quad (3)$$

where $c$ is the coupon and $F$ is the face amount of the bond.

On the other hand, with a sample of bonds price the parameters of the Nelson-Siegel and Svensson models can be estimated. The estimation is obtained by minimizing the objective function (4) with respect to $\boldsymbol{\theta} = (\beta_0, \beta_1, \beta_2, \lambda)$ parameters of Nelson-Siegel model or $\boldsymbol{\theta} = (\beta_0, \beta_1, \beta_2, \beta_3, \lambda_1, \lambda_2)$ parameters of Svensson model.

The objective function is given by the least square criterion with weighting factors proposed in [3]. These weighting factors allow using historical observations and it also reduces volatility through a stock measure:

$$\sum_{k=1}^{n} \frac{\left(Pr_k - \widetilde{Pr_k}\right)^2}{H_k (1 + ND_k)}, \quad (4)$$

where $Pr_k$ is the observed price for the bond $k$, $\widetilde{Pr_k}$ is the estimate price for the bond $k$ obtained by (3) as a function of $\boldsymbol{\theta}$, $ND_k$ is the number of days from the bond $k$ was traded and $H_k$ is the bid-ask spread:

$$H_k = \left| \frac{\sum_{i=1}^{m_s} OS_{k,i} fS_{k,i}}{fS_{k,total}} - \frac{\sum_{i=1}^{m_b} OB_{k,i} fB_{k,i}}{fB_{k,total}} \right|$$

where $OS_{k,i}$ (respectively $OB_{k,i}$) is the $i$-th sell (resp. buy) offer, $fS_{k,i}$ (resp. $fB_{k,i}$) is the facial amount of the $i$-th sell (resp. buy) offer, and $fS_{k,total}$ (resp. $fB_{k,total}$) is the total amount of facials sell (resp. buy) offers.

A set of constraints, similar to those used in [4] have been implemented with two goals, results economically feasible and speed in the optimization process.

The following constraints are for the Nelson-Siegel model:

$$0\% < \beta_0, \beta_2 < 25\%; -20\% < \beta_1 < 20\%; 1/300 < \lambda < 12; 0 < \beta_0 + \beta_1;$$

$$1/300 < \lambda < 12. \quad (5)$$

For the Svensson model constraints are:

$$0\% < \beta_0, \beta_2, \beta_3 < 25\%; -20\% < \beta_1 < 0\%; 1/300 < \lambda_1, \lambda_2 < 12;$$

$$0 < \beta_0 + \beta_1. \quad (6)$$

## 4 Optimization methods

In order to minimize (4), it is frequently used nonlinear regression methods based on Gauss-Newton, gradient descent or Marquardt iterative procedures [6]. However, it is well known that these procedures are suboptimal since they are based on local search; thus, they usually find a local minimum of the objective function. In order to avoid this suboptimality problem, in this article the following metaheuristics were used: Genetic algorithm [7], Ant colony [5], Particle swarm [13] and Simulated annealing [1]. These metaheuristics were programmed in R [24].

The results obtained with the metaheuristics were compared with the results of the Quasi-Newton algorithm BFGS [20] applied through an adaptive barrier method [15]. To implement these methods the built-in R [24] functions *constrOptim* and *optim* were used.

### 4.1 Genetic algorithm

Algorithm based on ideas of genetic evolution and biology [8], [18]. It starts with a population of solutions chosen randomly, in each iteration a new population is obtained from the previous one by pairing, mating and mutation. In our implementation, we use a population of $M = 100$ chromosomes, with a chromosomic representation based on a numerical vector of nonlinear regression parameters (4 parameters for Nelson-Siegel model, 6 parameters for Svensson model). Initial chromosomes are chosen at random satisfying the parameter constraints.

In this work fitness is the inverse of the cost function (4). Population matrix is ranked from best to worst. The best 50% are automatically kept as an elitist selection and the rest are replaced with the offspring generated by pairing, crossover and mutation.

For pairing, two chromosomes, the mother $\boldsymbol{\theta}^{mo} = (\theta_1^{mo}, \theta_2^{mo}, \dots, \theta_p^{mo})$ and the father $\boldsymbol{\theta}^{fa} = (\theta_1^{fa}, \theta_2^{fa}, \dots, \theta_p^{fa})$ are selected with probability:

$$p_i = \frac{M/2 - i + 1}{\sum_{m=1}^{M/2} m}.$$

Crossover is as follows: a crossing point is selected as the integer part of *up* plus one, with $u \sim U(0,1)$ and $p$ the number of parameters or variables in the regression model. Position $k$ of children is defined as

$$\theta_k^{ch1} = \theta_k^{mo} - \alpha(\theta_k^{mo} - \theta_k^{fa}), \theta_k^{ch2} = \theta_k^{fa} + \alpha(\theta_k^{mo} - \theta_k^{fa}), \alpha \sim U(0,1).$$

Children are defined by the exchange of variables at the right side of $k$:

$$\text{child}_1 = (\theta_1^{mo}, \dots, \theta_{k-1}^{mo}, \theta_k^{ch1}, \theta_{k+1}^{fa}, \dots, \theta_p^{fa})$$
$$\text{child}_2 = (\theta_1^{fa}, \dots, \theta_{k-1}^{fa}, \theta_k^{ch1}, \theta_{k+1}^{mo}, \dots, \theta_p^{mo}).$$

If $k=p$ then all positions at left are exchanged.

A mutation operator is performed over 1% of $(M-1) \times p$ positions in the population, excluding the best chromosome. Selected variable is replaced by a continuous random number (with uniform distribution) in the domain.

The algorithm stops if the standard deviation of fitness in population is less than 0.5 or if the maximum number of iterations (10,000) is attained.

### 4.2 Ant colony

Ant colony optimization (ACO) is a metaheuristic that takes its ideas from the way ants get food [5], [25], [26]. Usual ACO is usually designed for combinatorial optimization problems. In this study it is used the version for continuous domains presented in [26] since our case is rather continuous. The pheromones are used by means of an array that stores a number of solutions and new solutions are built sequentially using the information of the array.

ACO will construct sequentially a solution using a Gaussian kernel,

$$G^i(x) = \sum_{l=1}^{q} w_l g_l^i(x) = \sum_{l=1}^{q} w_l \frac{e^{-(x-\mu_l^i)^2/2(\sigma_l^i)^2}}{\sigma_l^i \sqrt{2\pi}}$$

where parameters are

$$\mu^i = (\mu_1^i, \dots, \mu_q^i) = (\theta_1^i, \dots, \theta_q^i), \sigma_l^i = \xi \sum_{h=1}^{q} \frac{|\theta_h^i - \theta_l^i|}{k-1},$$

$\xi$ being the evaporation rate of pheromone in ACO.

The weights are defined by:

$$w_l = \frac{e^{-(l-i)^2/2v^2 q^2}}{vq\sqrt{2\pi}}$$

where $l$ is the order of the $l$-th solution in decreasing order and $v$ is a user-defined parameter for speeding the convergence.

For constructing a solution $g_l^i$ is chosen with probability $p_l = w_l / \sum w_{l'}$. We take a random sample with distribution $g_l^i$ for completing a solution.

Best $q$ solutions are stored in $P = (S_1, \dots, S_q)$. In out implementation we used the following parameters: 2 ants, $q = 50$, $\xi = 0.4$ locality of the search and $v = 1.1$ speed of convergence.

### 4.3  Particle swarm

Based on the social behavior of some groups of animals [21], [30]. The performance of an individual is influenced by its best historical performance and the best overall performance of the group up to the present iteration.

In our implementation, each particle is a vector $\boldsymbol{\theta}$ in 4 dimensions (for the Nelson-Siegel model) or in 6 dimensions (for the Svensson model). We use a population $(\boldsymbol{\theta_1}, \ldots \boldsymbol{\theta_M})$ of $M = 47$ particles.

Let $\boldsymbol{\theta}^*(t)$ be the overall best particle and $\boldsymbol{\theta}_m^*(t)$ the best value for particle $m$ up to iteration $t$. Then next position of particle $m$ in iteration $t+1$ is:
$$\theta_m(t+1) = \theta_m(t) + v_m(t+1)$$
where $\boldsymbol{\theta_m}(t)$ is its position in iteration $t$ and
$$v_m(t+1) = w(t)v_m(t) + \lambda_1 r_1[\theta^*(t) - \theta_m(t)] + \lambda_2 r_2[\theta_m^*(t) - \theta_m(t)]$$
is the velocity vector, that defines the direction of the particle in the new iteration, with
$$w(t) = w_{max} - (w_{max} - w_{min})\frac{t}{T_{max}}.$$
Here, $\lambda_1$ is a cognitive parameter and $\lambda_2$ is a social parameter; $r_1, r_2 \sim U[0,1]$ are random numbers. We suppose that velocity is bounded $|v_{mj}(t)| \leq v_{max}$ so the particles do not diverge, $w_{max}$ and $w_{min}$ are bounding parameters and $T_{max}$ is a maximum number of iterations. We iterate until $\boldsymbol{\theta}_m^*(t)$ does not change or iterations reach $T_{max}$.

Taking into account the recommendations made by [9], $w = -0.1832$, $\lambda_1 = 0.5287$ as cognitive parameter and $\lambda_2 = 3.1913$ as social parameter.

### 4.4  Simulated annealing

Based on the physical process named annealing, which takes a solid to a high temperature and then let it cool very slowly in order to get a more resistant and pure state of the solid [1], [16], [29]. Also, it uses the Metropolis criterion of acceptance whose purpose is to get out of local minimum zone [16], [27], [29]; this criterion accepts better states of the problem, but may also accept a worse state with a certain probability, that decreases as the temperature cools down.

It is well known that, from a Markov chain modeling, simulated annealing converges asymptotically to the global optimum under some conditions [1]. Basic conditions of the Markov chains are reversibility, connectedness and length of the chains.

In this paper it is used the version named very fast simulated reannealing [11] which allows to work with restrictions.

Let $\boldsymbol{\theta}$ be a state of the problem, that is, a set of 4 or 6 nonlinear regression parameters, depending on dealing with the Nelson-Siegel or the Svensson model, respectively. A new state $\boldsymbol{\theta}'$ will be defined by components generated as
$$\theta_i' = \theta_i + \lambda_i \left(\theta_{max_i} - \theta_{min_i}\right)$$
where $\lambda_i \in [-1,1]$ and $\theta_{max_i}, \theta_{min_i}$ are bounds of the $i$-th parameter. Let $T$ be the simulated annealing temperature, we use $\lambda_i$ distributed with

$$g_T(\lambda_i) = \frac{1}{2(|\lambda_i| + T)\ln(1 + 1/T)}$$

where $\lambda_i$ is generated as $\lambda_i + sgn(u - 0.5)T[(1 + 1/T)^{|2u-1|} - 1]$, and $u \sim U[0,1]$.

The size of the Markov chain was established in 100, and the temperature is updated with the factor 0.95, that is $T_{k+1} = 0.95 T_k$.

For estimating the initial temperature, we follow [1]. Given a value $\chi_0 \approx 0.95$ that represents the fact that, at the beginning, almost 95% of new states that worsen the objective function $F$ in equation (4) will be accepted in the Metropolis rule. Then, is we make 1000 blank iterations let $m_1$ be the number of times that $F$ decreases and $m_2$ the number of times that $F$ increases; if $\overline{\Delta F}^+$ is the average in $F$ differences for those blank iterations that increase the value of $f$, then $T_0$ is estimated with

$$T_0 = \overline{\Delta F}^+ / \ln\left(\frac{m_2}{m_2 \chi_0 - m_1(1 - \chi_0)}\right).$$

Metropolis rule works as follows: a new state is accepted if $f$ decreases, or it is accepted with probability

$$\exp(-\Delta F/T),$$

where $\Delta F = F(\boldsymbol{\theta}') - F(\boldsymbol{\theta})$.

The iterations stop when $T \approx 0$, or a maximum number of iterations is reached, or after one complete Markov chain there are no improvements in the cost function.

### 4.5 An adaptive barrier with a BFGS quasi-Newton algorithm

BFGS algorithm is a local search method [20] where the search is given by a modified Newton direction.

Let $\boldsymbol{\theta}(t)$ be the current state of the problem, the new state is defined by a vector direction $\boldsymbol{p}(t)$ as in several descent methods:

$$\boldsymbol{\theta}(t + 1) = \boldsymbol{\theta}(t) + \alpha_t \boldsymbol{p}(t)$$

such that $F(\boldsymbol{\theta}(t + 1)) \leq F(\boldsymbol{\theta}(t))$, where $\alpha_t \in \mathbb{R}$, $\alpha_t = \arg\min_{(\alpha > 0)} F(\boldsymbol{\theta}(t) + \alpha \boldsymbol{p}(t))$.

With a second order Taylor approximation for $F(\boldsymbol{\theta}(t) + \boldsymbol{p}(t))$ it is obtained

$$\boldsymbol{p}(t) = -\left(\nabla^2 F(\boldsymbol{\theta}(t))\right)^{-1} \nabla F(\boldsymbol{\theta}(t))$$

supposing $\nabla^2 F(\boldsymbol{\theta}(t))$ is positive definite. In BFGS algorithm [20], Hessian is replaced by an approximation calculated in each iteration:

$$H_{t+1} = (I - \frac{\boldsymbol{z}(t)\boldsymbol{y}'(t)}{\boldsymbol{y}'(t)\boldsymbol{z}(t)})H_t(I - \frac{\boldsymbol{y}(t)\boldsymbol{z}'(t)}{\boldsymbol{y}'(t)\boldsymbol{z}(t)}) + \frac{\boldsymbol{z}(t)\boldsymbol{z}'(t)}{\boldsymbol{y}'(t)\boldsymbol{z}(t)}$$
$$\boldsymbol{z}(t) = \boldsymbol{\theta}(t + 1) - \boldsymbol{\theta}(t), \quad \boldsymbol{y}(t) = \nabla F(\boldsymbol{\theta}(t + 1)) - \nabla F(\boldsymbol{\theta}(t)).$$

In order to satisfy the constraints in the Nelson-Siegel and Svensson models, we have used an adaptive barrier method, that transforms the minimization problem

$$\min_{\boldsymbol{\theta}} F(\boldsymbol{\theta}) \text{ subject to } L_j(\boldsymbol{\theta}) = \boldsymbol{u}_j' \boldsymbol{\theta} - c_j \geq 0, j = 1, \ldots, p$$

into

$$\min_{\boldsymbol{\theta}} F(\boldsymbol{\theta}) - \mu \sum_{i=1}^{p} [L_j(\boldsymbol{\vartheta}_k) \ln L_j(\boldsymbol{\theta}) - \boldsymbol{u}_j' \boldsymbol{\theta}],$$

where $F(\boldsymbol{\theta})$ has been added with a so-called logarithmic barrier that considers the regression constraints, $\vartheta_k$ is an interior point of the feasible region. Parameter $\mu$ tends to 0 with the goal to neglect more and more the barrier [15].

The objective function depends on the vector $\boldsymbol{\theta}$ which has to satisfy (5) or (6), so that the constrained optimization problem is changed into an unconstrained problem, and an adaptive barrier method is used [17]. In this case a logarithmic barrier is added to the objective function in order to handle the constraints (5) or (6). If the minimum lies on the boundary the barrier will not allow to reach it, to deal with this the logarithmic barrier has a component that changes in each iteration [15].

In the minimization of the barrier method, the BFGS procedure is used.

## 5  Results

For each method a multistart strategy [30] of size 2,000 was made. The way of comparison is as follows: the best objective function value for the metaheuristics is the expected value from their multistart, in the case of the adaptive barrier the best objective function value is the minimum value that was achieved from its multistart.

Tables 1 and 2 contain a summary of the results for the Nelson-Siegel and Svensson models, respectively. The following values are reported: the objective function value, the coefficient of variation information taken from the multistart, the goodness of fit and the average time of running the R-code measured in seconds.

**Table 1.** Summary metaheuristics performance in estimating the Nelson-Siegel model.

| Algorithm | Objective function value | Coefficient of variation | Goodness of fit | Average time (m) |
|---|---|---|---|---|
| Particle swarm | 441.5018 | <1% | 0.003345% | 00:22 |
| Simulated ann. | 441.5034 | <1% | 0.003353% | 00:28 |
| Adaptive barr. | 441.5243 | 240% | 0.003354% | – |
| Genetic alg. | 1,206.0571 | 18% | 0.066502% | 01:16 |
| Ant colony | 1,207.6136 | 36% | 0.066541% | 00:18 |

The results for the Nelson-Siegel model are shown in the Table 1. Two metaheuristics got better results than the adaptive barrier method, namely, Particle swarm and Simulated annealing. Their coefficients of variation are almost zero indicating that the same results are obtained almost every time the functions are run. The average time is approximately 20 seconds.

Figure 1 shows graphically the yield curves obtained with the four metaheuristics for Nelson-Siegel model.

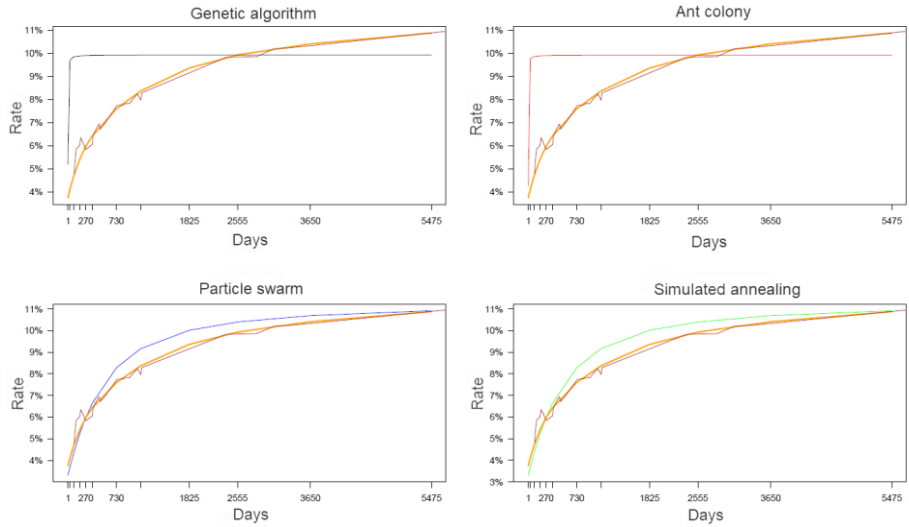

**Fig. 1.** Yield curve and estimated yield curve with Nelson-Siegel model for March 17, 2015.

**Table 2.** Summary metaheuristics performance in estimating the Svensson model.

| Algorithm | Objective function value | Coefficient of variation | Goodness of fit | Average time (m) |
|---|---|---|---|---|
| Particle swarm | 251.5805 | <1% | 0.012147% | 00:43 |
| Simulated ann. | 251.6899 | <1% | 0.012550% | 00:46 |
| Ant colony | 254.6444 | 84% | 0.012345% | 01:32 |
| Adaptive barr. | 441.6267 | 317% | 0.003164% | – |
| Genetic alg. | 1,138.3852 | 12% | 0.052407% | 00:13 |

In the case of the Svensson model (see Table 2) three metaheuristics had better performance than the adaptive barrier: Particle swarm, Simulated annealing and Ant colony. But we highlight Particle swarm and Simulated annealing which have a coefficient of variation almost zero and an average time of 40 seconds.

In Figure 2 are shown graphically the yield curves obtained for the Svensson model.

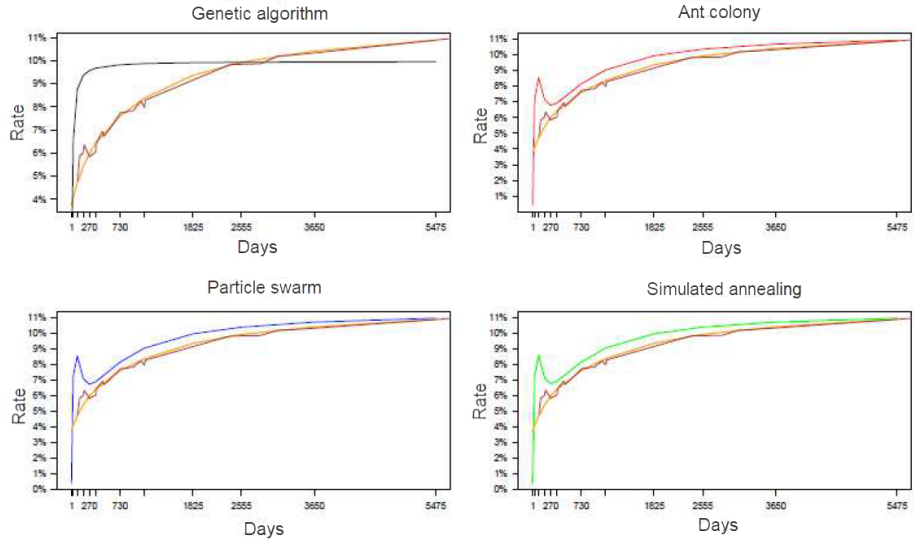

**Fig. 2.** Yield curve and estimated yield curve with Svensson model for March 17, 2015.

## 6      Concluding remarks and further research

Two metaheuristics were better in both models, Particle swarm and Simulated annealing. These metaheuristics besides of having the best results, their algorithms are easy to implement, the execution time is acceptable, and the outcomes are very stable.

Therefore, Particle swarm and Simulated annealing are recommended for getting the parameters of the Nelson-Siegel and Svensson models.

For future research it is suggested to repeat this study with other sets of sample data so as to confirm the results obtained so far. For another financial data set, similar restrictions for (5) or (6) which are adjusted to the Costa Rican market characteristics, should be determined. Moreover, parameters tuning can be improved with a factorial design that could suggest better choices. Finally, a review of the configuration used in Genetic algorithm and Ant colony could also be made in order to obtain satisfactory parameters that may make compete these metaheuristics with the better ones. Also, we will perform further studies with simulated data and controlled parameters, and the use of benchmark data will also be considered.